# Business fluctuations in a credit-network economy

Domenico Delli Gatti,[a*] Mauro Gallegati,[b] Bruce Greenwald,[c]
Alberto Russo,[b] Joseph E. Stiglitz,[c]

[a] Catholic University, Milan, Italy
[b] Politechnic University of Marche, Ancona, Italy
[c] Columbia University, New York, USA

**Abstract**

We model a network economy with three sectors: downstream firms, upstream firms, and banks. Agents are linked by productive and credit relationships so that the behavior of one agent influences the behavior of the others through network connections. Credit interlinkages among agents are a source of bankruptcy diffusion: in fact, failure of fulfilling debt commitments would lead to bankruptcy chains. All in all, the bankruptcy in one sector can diffuse to other sectors through linkages creating a vicious cycle and bankruptcy avalanches in the network economy. Our analysis show how the choices of credit supply by both banks and firms are interrelated. While the initial impact of monetary policy is on bank behaviour, we show the interactive play between the choices made by banks, the choices made by firms in their role as providers of credit, and the choices made by firms in their role as producers.

\* *Corresponding author*: domenico.delligatti@unicatt.it (D. Delli Gatti).

# 1. Introduction

We model a network economy with an *inside credit* (*commercial credit*) between firms of different productive sectors, and an *outside credit* (*bank credit*), which lends credit to industrial sectors (Stiglitz, Greenwald, 2003). Failure of fulfilling debt commitments could lead to *bankruptcy chains*. If debt commitments are not fulfilled, *bad debt* increases and the rate of interest may increase as well. The interest rate increase leads to more bankruptcies: *"the high rate of bankruptcy is a consequence of the high interest rate as much as a consequence of it"* (Stiglitz, Greenwald, 2003: 145).

If future is uncertain, what probability is there that the contract will be fulfilled? Are there major consequences of a small shock? This paper will show in what way a business cycle is linked to firm bankruptcies and how a domino effect (a *crescendo* in terms of bankruptcy) can arise (and as such be avoided). At the same time, the main facts of firm demography (like power law distribution of size and Laplace growth rates of firms) emerge endogenously (and are more resistant to external shocks.) As opposed to previous models of business cycles, in our model, avalanches are due to the interdependence of the output of a firm on supply and payments from other firms. Similarly to the self-organized criticality inventory cycle proposed by Bak *et al*. (1993), in our model, small shocks fail to cancel at the aggregate level so that large economic fluctuations can occur even without aggregate shocks. According to Bak *et al*. (1993), we model productive relationships among agents as local interactions (with a different network topology). In addition, we consider firm-bank and firm-firm credit relationships that can amplify the effects of small idiosyncratic shocks on the aggregate.

Heterogeneous agent interaction has a second major implication. We will see that the structure of aggregate behaviour (*macro*) actually emerges from the interaction among the agents (*micro*). In other words, statistical regularities emerge as a self-organised process at the aggregate level. Complex patterns of interacting individual behaviour may generate certain regularity at the aggregate level (Delli Gatti *et al*., 2005). The idea of representing a society by one exemplar (a *representative agent*) denies the fact that the organizational features of the economy play a crucial role in explaining what happens at the aggregate level (Kirman, 1992). The model is presented in section 2. The model simulation and the discussion of results are shown in section 3. Section 4 concludes.

# 2. The model

Consider a sequential economy ($t=1,2,…,T$) populated by a multitude of heterogeneous agents belonging to three different sectors: a downstream sector with a number $I$ of firms (labeled by the index $i=1,2,…,I$), an upstream sector with $J$ firms ($j = 1,2,…,J$) and a banking sector with $Z$ banks ($z=1,2,…,Z$). Downstream firms produce and sell final goods (at a stochastic price). The production of the final goods requires two productive inputs: labor and intermediate goods. The intermediate goods are produced "on demand" by upstream firms with a technology that requires only labor as input. Upstream firms sell intermediate goods to downstream firms by means of a commercial credit contract. Firms obtain credit from banks to finance the wage bill. Finally, banks are linked in the interbanking market in order to solve liquidity crises.

The network structure of the economy is the following (see figure 1): each downstream firm $i$ is linked to upstream firms $j$ and $j+1$; the downstream firm $i+1$ is linked to upstream firms $j+1$ and $j+2$, and so on.[1] Each firm is linked to a bank: the downstream firm $i$ and the upstream firm $j$ are linked to the bank $z$, and so on. Finally, each bank is linked to two neighboring banks: bank z is linked to banks $z-1$ and $z+1$. All in all, downstream firms, upstream firms and banks are linked by productive and credit relationships so that the behavior of one agent influences the behavior of the others through network connections.

The production function of downstream firms is given by the equation: $Y_{it}=\phi A^{\beta}_{it-1}$, where $\phi>1$, $0<\beta<1$, and $A_{it}$ is the net worth of the firm $i$ at time $t$. The production of final goods needs two inputs. Hence, downstream firms have the following labor and intermediate goods requirement functions: $N_{it}=\delta_d Y_{it}$, $Q_{it}=\gamma Y_{it}$, with $\delta_d>0$ and $\gamma>0$. Final goods are sold at a stochastic price $u_{it}$.[2] Upstream firms produce intermediate goods required by the downstream sector by means of a technology employing only labor as input: $Q_{jt}= \delta_u N_{jt}$, where $\delta_u>0$ and, for assumption, $Q_{jt}=Q_{i-1t}/2+Q_{it}/2$. The output of the upstream sector is sold to downstream firms at a fixed price $p$.

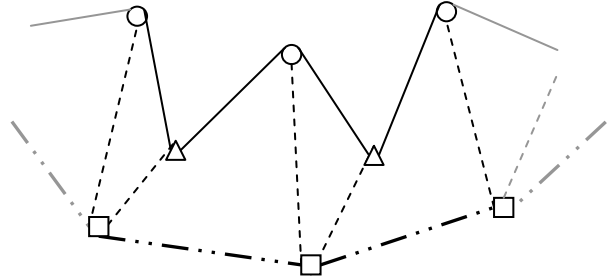

**Figure 1**. The network structure of the economy:
Downstream firms (○), upstream firms (Δ), and banks (□).

Downstream firms sign a commercial credit contract with upstream firms so that they obtain the ordered quantity of the intermediate goods from upstream firms and pay back for it in the next period. Upstream firms charge an interest rate $r^u$, constant along time and uniform across firms, on commercial loans. The wage bill is financed by bank loans with an interest rate $r^{bx}_{it}$ ($x=d$ for downstream, $x=u$ for upstream firms).[3] We assume that there is an infinite supply of labor at wage $w$.

Downstream firms' profit is equal to: $\pi_{it}=u_{it}Y_{it} - (1+r^{bd}_{it})wN_{it} - (1+r^u)pQ_{jt}$, where $r^{bd}_{it}$ is the interest rate on the bank loan. Upstream firms' profit is equal to: $\pi_{jt}=p(1+r^u)Q_{jt} - (1+r^{bu}_{jt})wN_{jt}$, where $r^{bu}_{jt}$ is the interest rate on the bank loan. At the end of the period, firms' net worth is equal to $A_{xt}=A_{xt-1}+\pi_{xt}$, and a firm goes bankrupt if $A_{xt}<0$ ($x=i$ for downstream; $x=j$ for upstream). We assume a very simplified mechanism of agents' entry-exit: bankrupted

---

[1] Accordingly, the downstream firm *I* is linked to upstream firms *J* and 1.
[2] That is a random variable that, for the sake of simplicity, is uniformly distributed in the interval (0,2).
[3] See below for the interest rate setting.

firms are replaced with new entrants on the basis of a one-to-one replacement. The same holds for banks.[4]

The commercial credit relationship between downstream and upstream firms is a source of bankruptcy diffusion across the network. In fact, if a downstream firm goes bankrupt then it cannot refund the debt to the upstream ones increasing their probability of bankruptcy (due to the decrease of their net worth). The deterioration of the financial conditions of upstream firms jointly with the failure of downstream firms can cause bankruptcies even in the banking sector, given that bankrupted firms cannot refund loans to banks. All in all, bankruptcy in one sector can spread to other sectors through connections creating a vicious cycle and bankruptcy avalanches in the network economy, as we will illustrate in section 3.

Banks lend funds to downstream and upstream firms in order to finance their production process, that is their wage bills. The total amount of credit each bank can supply to firms is equal to $L^s_{zt}=E_{zt}/\alpha$, where $E_{zt}$ is the level of bank $z$'s net worth at time $t$ and $\alpha<1$ is a prudential target set by a regulatory authority that banks have to respect. In other words, the level of credit supply is more than proportional to banks' net worth and limited to a maximum depending on the coefficient $\alpha$. In each period banks collect the demand for credit from firms, $L^d_{zt}$. Given that the supply of credit is limited to $L^s_{zt}$, firms can be rationed for an amount proportional to $L^s_{zt}/L^d_{zt}$ and revise their productive plans on the basis of the effective credit obtained from banks.

For the sake of simplicity, we assume that deposits $D_{zt}$ are a residual variable, given the balance sheet of banks. The interest rate paid by banks on deposits is $r^d$, constant along time and uniform across banks. The interest rate on bank loans is: $r^{bx}_{yt}=k/(A_{yt}/\tilde{A}_{xt})^k$, where $0<k<1$, and $\tilde{A}_{xt}$ is the median of the firms' net worth distribution at period $t$, computed separately for upstream and downstream firms ($x=d$ and $y=i$ for downstream firms, $x=u$ and $y=j$ for upstream firms). Accordingly, the higher the financial fragility of firms the higher the interest rate charged on bank loans. Banks' profit is equal to: $\pi^b_{zt}=r^{bd}_{it}wN_{it} + r^{bu}_{jt}wN_{jt} - r^d D_{zt}$. The level of the banks' net worth evolves according to the following equation: $E_{zt}=E_{zt-1} + \pi^b_{zt} - BD_{zt} + IB_{zt}$, where $BD_{zt}$ is "bad debt" for bank $z$, due to the insolvencies produced by upstream and/or upstream firms' failures, and $IB_{zt}$ is the result of liquidity adjustments on the interbank market.

Banks can lend or borrow funds on the interbank market in order to solve liquidity crises or to invest resources in excess with respect to firms' demand for credit. We assume that the rate of interest on the interbank market, $r^{bb}$, is constant along time and uniform across banks. As said above, bank $z$ interacts only with neighbors $z-1$ and $z+1$. Accordingly, in period $t$, each bank can be a fund lender or borrower with respect to the neighbor banks:

$$IB_{zt} = \begin{cases} (1+r^{bb})IB^l_{zt-1} + IB^b_{zt} & (i) \\ (1+r^{bb})IB^l_{zt-1} - IB^l_{zt} & (ii) \\ -(1+r^{bb})IB^b_{zt-1} + IB^b_{zt} & (iii) \\ -(1+r^{bb})IB^b_{zt-1} - IB^l_{zt} & (iv) \end{cases}$$

---

[4] Accordingly, the total number of agents in the economy is constant along time. New agents are endowed with an initial amount of net worth equal to that choose for initial conditions (that is, the value of agents' variables at time $t=1$).

where $l$ and $b$ indexes mean lending and borrowing, respectively. Let us explain the above four scenarios: (i) bank $z$ lent funds in $t–1$ (and receives reimbursement with interest in $t$) and borrows funds in $t$; (ii) bank $z$ lent funds in $t–1$ and lends funds in $t$; (iii) bank $z$ borrowed funds in $t–1$ (and refunds the loan with interest in $t$) and borrows funds in $t$; (iv) bank $z$ borrowed funds in $t–1$ and lends funds in $t$. Even though banks operate in the interbank market to manage their liquidity and, in particular, to avoid defaults due to firms' bad debts, the interbanking activity can be an additional source of spread of bankruptcies (Iori *et al*., 2004). In fact, a borrowing bank may go bankrupt in a given period generating insolvency for the lending bank. The financial conditions of this last bank can degenerate, resulting in another failure. So, as in the case of firm-firm and firm-bank credit relationships, interbank connections can generate avalanches of defaults, reinforcing the vicious circle of spreading of bankruptcies across the network economy.

## 3. Simulations

We simulate the network economy described above with $I=J=Z=250$ agents in each sector and analyze the model simulation from period $t=1$ to $T=1000$. At the beginning of the simulation, agents (firms and banks) start with an amount of net worth equal to 100. The wage is fixed at level $w = 1$. The same holds for the intermediate goods price, $p = 1$; the interest rate on commercial credit, $r^u = 0.05$; the interest rate on deposits, $r^d=0.01$; the interest rate on interbank loans, $r^{bb} = 0.01$.[5]

Figure 2 describes the fluctuating behavior of the average production of downstream firms. Nevertheless, agents start with the same initial condition. Simulations show that a significant heterogeneity emerges in agent variables. For example, the final distribution (at period $T=1000$) of the net worth of downstream firms is highly asymmetric and skewed to the right (Axtell, 2001). The same result holds for upstream firms and banks. In addition, figure 3 shows the difference between the firm size distribution that emerges from simulations and a normal one.[6] The model also generates a tent-shaped (Laplace) distribution of firm growth rates (Stanley et al., 1996; Bottazzi, Secchi, 2003), as shown in Figure 4.

---

[5] The parameter setting of the simulation is the following: $\phi$=2.5, $\beta$=0.9 (production function parameters: downstream firms); $\gamma$=0.5 (intermediate goods requirement function coefficient: downstream firms); $\delta_d$=0.5 (labor requirement function coefficient: downstream firms); $\delta_u$=1 (labor requirement function coefficient: upstream firms); $k$=0.1 (firm-bank interest rate function parameter); $\alpha$=0.85 (prudential target for credit supply: banks)

[6] The Bera-Jarque test refuses the null hypothesis of normality of the agent size distribution at 1% significance level.

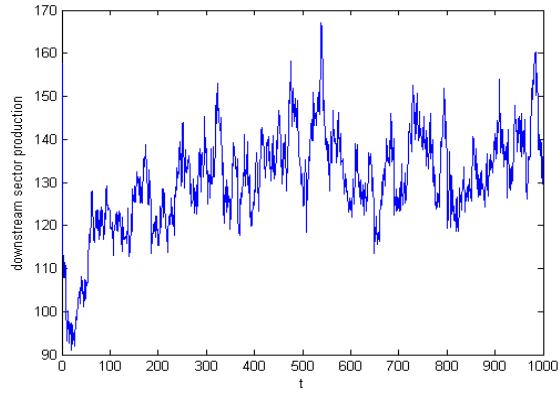

**Figure 2**. Downstream sector average production

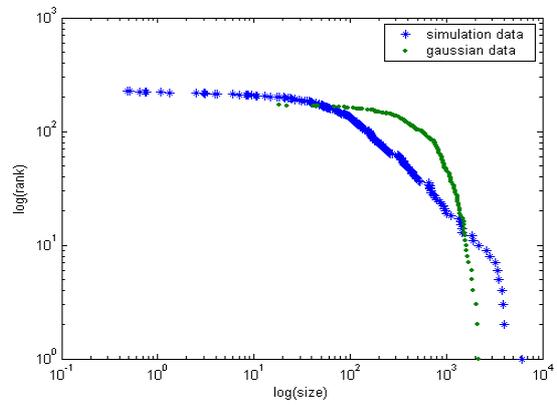

**Figure 3**. Downstream firm size distribution (net worth)

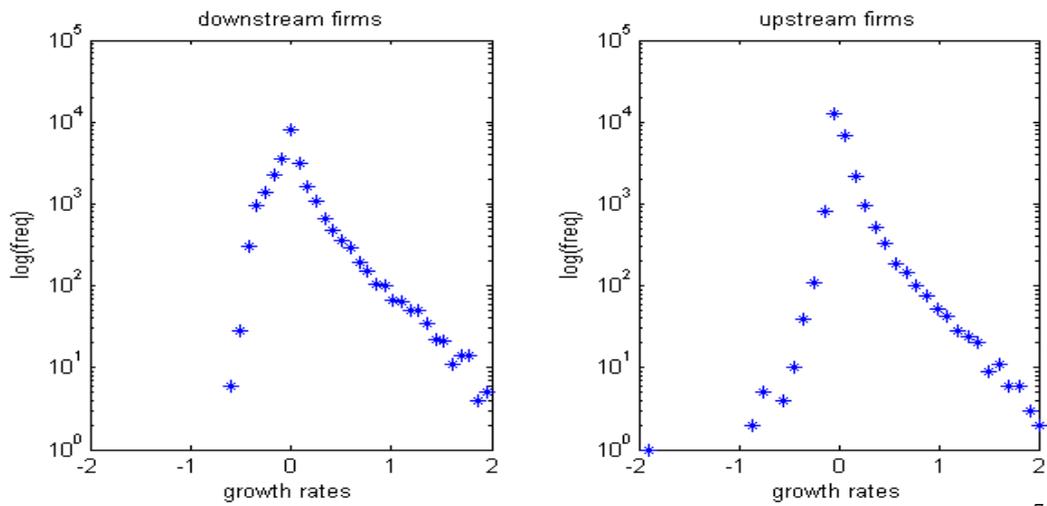

**Figure 4**. The Laplace (double exponential) distribution of firm growth rates.[7]

---

[7] Pooled data from the last 100 simulation periods.

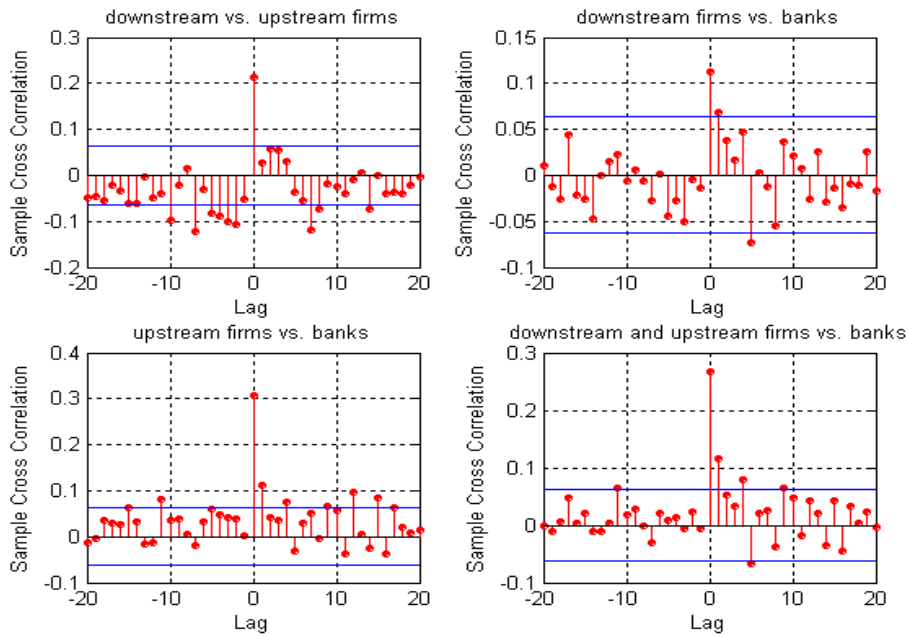

**Figure 5**. Cross correlation of bankruptcies in different sectors

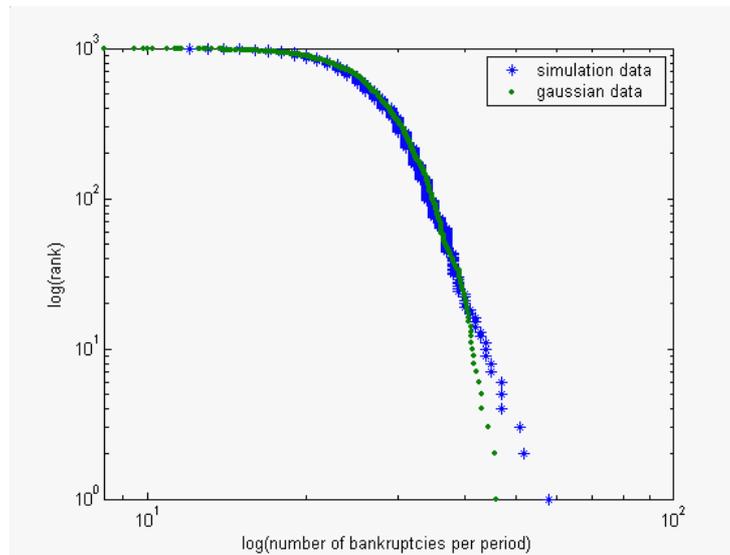

**Figure 6**. Avalanches size distribution

A central feature of the network economy described in our model is related to the spread of bankruptcies through the productive and credit linkages among agents. If a downstream firm goes bankrupt it cannot refund the debt to the upstream firms it is linked to, decreasing their net worth, and increasing their probability of bankruptcy. The weakening of the financial conditions of upstream firms jointly with the failure of downstream firms can cause bankruptcies even in the banking sector, given that bankrupted firms cannot refund loans to banks. All in all, bankruptcy in one sector can spread to other sectors through network connections creating a vicious cycle and bankruptcy avalanches in the economy. Figure 5 shows that the number of bankruptcies per period in a sector is correlated to the number of bankruptcies in other sectors. In addition, figure 6 shows that the distribution of the total number of bankruptcies per period has a fatter tail with respect to a normal distribution,[8] so that "rare events" are more probable in this setting than in an economy characterized by normal distributions.

## 4. Conclusive remarks

The idea of a snowball phenomenon in which the bankruptcy of one firm somehow affects the financial conditions of the other firms forcing the most vulnerable among them into bankruptcy is not new in economic literature. In this paper the idea is framed in the context of a network of economy in which firms are linked by production and commercial relationship. The bankruptcy of a firm makes banks less willing to extend loans to other firms. A reduction of credit to a firm or an increase in the interest rate charged affects the willingness and ability of that firm to supply commercial credit to its customers. As its customers are adversely affected, they transmit the contraction of credit on to their customers, and so forth around the economy. The higher interest rates charged by suppliers imply higher bankruptcy probabilities for these firms and this in turn induces banks to cut back on their credit.

Our analysis has shown how the decision to provide credit by both banks and firms are interrelated. While the initial impact of monetary policy is on bank behaviour, we showed the interactive play between the choices made by banks, the choices made by firms in their role as providers of credit, and the choices made by firms in their role as producers.

---

[8] The Bera-Jarque test refuses the null hypothesis of normality of the bankruptcies distribution (1% significance level).